\documentclass[12pt]{article}
\usepackage{amsmath,amssymb}

\usepackage{graphicx}
\usepackage{wrapfig}

\usepackage{hyperref}
\usepackage{cite}

\def\beq{\begin{eqnarray}}
\def\eeq{\end{eqnarray}}

\def\o{\over}

\newcommand{\Tr}{\,\mathrm{Tr}\,}            

\newcommand{\be}{\begin{equation}}
\newcommand{\ee}{\end{equation}}
\newcommand{\bea}{\begin{eqnarray}}
\newcommand{\eea}{\end{eqnarray}}
\newcommand{\bg}{\begin{gather}}
\newcommand{\eg}{\end{gather}}
\newcommand{\bseq}{\begin{subequations}}
\newcommand{\eseq}{\end{subequations}}

\renewcommand{\ln}{\mathop{\rm ln}\nolimits}

\def\be{\begin{eqnarray}}
\def\ee{\end{eqnarray}}
\def\lb{\label}

\begin{document}

\title{\textbf{Short-distance regularity of Green's function and UV divergences in entanglement entropy }}

\vspace{1cm}
\author{ \textbf{Dmitry Nesterov$^{\star}$ and
 Sergey N. Solodukhin$^\sharp$ }} 

\date{}
\maketitle

\begin{center}
  \hspace{-0mm}
  \emph{$^{\star}$ $^{\sharp}$ Laboratoire de Math\'ematiques et Physique Th\'eorique }\\
  \emph{Universit\'e Fran\c cois-Rabelais Tours F\'ed\'eration Denis Poisson - CNRS, }\\
  \emph{Parc de Grandmont, 37200 Tours, France} \\
  \emph{ and }\\
  \emph{$^{\star}$ Theory Department, Lebedev Physical Institute,
      \\ Leninsky Prospect 53, Moscow, Russia, 119991.}
\end{center}

{\vspace{-11cm}
\begin{flushright}
\end{flushright}
\vspace{11cm}
}



\begin{abstract}
\noindent { Reformulating our recent result (arXiv:1007.1246 [hep-th]) in coordinate space we point out
that no matter how regular is short-distance behavior of Green's function
the entanglement entropy in the corresponding quantum field theory is always UV divergent. In  particular, we discuss a recent example by Padmanabhan (arXiv:1007.5066 [gr-qc]) of a regular Green's function and show that provided this function arises in a field theory the entanglement entropy in this theory is UV divergent and calculate the leading divergent term.}
\end{abstract}

\vskip 3 cm
\noindent
\rule{9.7 cm}{.5 pt}\\
\noindent 
~~~$^{\star}$ {\footnotesize e-mail: nesterov@lpi.ru, Dmitry.Nesterov@lmpt.univ-tours.fr}\\
\noindent ~~~$^{\sharp}$ {\footnotesize e-mail: Sergey.Solodukhin@lmpt.univ-tours.fr}


\newpage

\section{ Introduction}
\setcounter{equation}0
Entanglement entropy  \cite{BS}, \cite{Srednicki:1993im} remains a fascinating subject of current research (for a recent review see \cite{Casini:2009sr}). It is defined by tracing  degrees of freedom residing inside a surface $\Sigma$ and, to some degree, measures the short-distance correlations across the surface. Its geometrical feature (the proportionality to the area $A$ of the surface)  makes it
a very attractive candidate to provide a statistical explication to the gravitational entropy associated to horizons. A major difficulty on this way, however, is the fact that
entanglement entropy calculated for a quantum free field is UV divergent.  In $d$ space-time dimensions one has that
\be
S_{ent}\sim {A\o \epsilon^{d-2}}~~,
\lb{1}
\ee
where $\epsilon$ is an UV cut-off.
A simple (and perhaps somewhat naive) way to understand the origin of the divergence in the entropy is to relate it to the short-distance divergence of the
2-point function, in $d$ space-time dimensions one has in a standard field theory
\be
<\phi(x),\phi(y)>=G(x,y)= {\Omega_d\o |x-y|^{d-2}}~~,
\lb{2}
\ee
where $\Omega_d={\Gamma ({d-2\o 2})\o 4\pi^{d\o 2}}$ and $G(x,y)$ is Green's function. At first sight this relation seems natural:
two sub-systems separated by surface $\Sigma$ know about each other due to the short-distance correlations that exist between the modes residing on different sides of the surface. As a result of this correlation the entropy is non-vanishing and is determined by geometry of the surface, to leading order by the area. Since the short-distance correlations are divergent, as in (\ref{2}), this divergence seems to manifest in the UV divergences in the entropy (\ref{1}).

Therefore, one may  think that there is a one-to-one correspondence between divergences in (\ref{1}) and (\ref{2}) so that one may
expect that in a theory in which 2-point functions are regular in the coincidence limit the entanglement entropy would be  automatically UV finite. A simple example of this sort is the following modification of (\ref{2})
\be
G_L(x,y)={\Omega_d\o ((x-y)^2+L^2)^{d-2\o 2}}~~,
\lb{3}
\ee
where the short-distance divergences are now regularized by parameter $L$. This example was recently considered by Padmanabhan \cite{Padmanabhan:2010wg}
who argued that in a theory with Green's function (\ref{3}) entanglement entropy is UV finite. In his approach the parameter $L$ incorporates some fundamental, possibly stringy, effects so that $G_L$, as it appears in  \cite{Padmanabhan:2010wg}, is seemingly not a field theoretical Green's function.
Nevertheless, since (\ref{3}) may well appear in some field theory, one may have impression from reading \cite{Padmanabhan:2010wg} that in the field theory  the regularity of Green's function in the coincidence limit implies the UV finiteness of entanglement entropy. The latter is not the case as we show in this note.

In a recent paper \cite{Nesterov:2010yi} we have proved a sort of ``no-go theorem'' by showing that no matter how well is UV behavior of the propagator of a quantum field theory the entanglement entropy calculated in this theory is always UV divergent. We have showed this by using momentum representation of the heat kernel for a  quantum field satisfying a rather general (Lorentz invariant or non-Lorentz invariant) field equation.
Obviously a field theory, in which (\ref{3}) appears as Green's function, belongs to the class of theories we have considered and hence entanglement entropy in this theory is still UV divergent even though the short-distance correlations in (\ref{3}) are regular.

In this note we first reformulate the statement made in \cite{Nesterov:2010yi} in terms of Green's function and then discuss a theory with Green's function of the type (\ref{3}).  For simplicity we only consider  Lorentz invariant field theories.

\section{ Heat kernel, Green's function and entanglement \\ entropy}
\setcounter{equation}0
We consider a  quantum field  that satisfies a rather general Lorentz invariant field equation
\begin{equation} \lb{F}
  {\cal D}\phi=F(\Box)\phi=0~,
 \end{equation}
where $F(\Box)$ is  an arbitrary function of the Laplace operator $\Box=-\partial_\mu\partial^\mu$. Many important quantities that characterize the quantum field can be expressed in terms of the heat kernel $K(s,X,X')=<X|e^{-s{\cal D}}|X'>$ (for a standard review on the heat kernel method see \cite{Vassilevich:2003xt}). The latter is defined as a solution  to the heat  equation
 \begin{equation}\label{K}
  \left\{
  \begin{array}{l}
    (\partial_s+{\cal D}) \,K(s,X,Y)=0 \,, \\
    K(s{=}0,X,Y)=\delta(X,Y) \,.
  \end{array}
    \right.
 \end{equation}
 In particular,  the effective action is defined as
 \begin{equation}
  W=-{1\o 2}\int_{\epsilon^2}^\infty {ds\o s}\Tr K (s)\lb{W}\,,
 \end{equation}
where parameter $\epsilon$ is an UV cutoff.
In flat spacetime one can use the Fourier transform  in order to solve the heat equation (\ref{K}). In $d$ spacetime dimensions one has
 \begin{equation}
  K(s,X,Y)={1\over (2\pi)^d}\int d^dp \,e^{ip_\mu(X^\mu{-}Y^\mu)}~e^{-sF(p^2)}\,.
 \lb{2-K}
 \end{equation}
 Note that we consider Euclidean theory so that $p^2\geq 0$.
The Green's function is a solution to the field equation with a delta-like source
\be
{\cal D} \, G(X,Y)=\delta(X,Y)
\lb{G}
\ee
and can be expressed in terms of the heat kernel as follows
\be
G(X,Y)=\int_0^\infty ds \, K(s,X,Y)~~.
\lb{GK}
\ee
Obviously, Green's function can be represented in terms of the Fourier transform in a manner similar to (\ref{2-K}),
\be
G(X,Y)={1\over (2\pi)^d}\int d^dp ~e^{ip_\mu(X^\mu{-}Y^\mu)}~ G(p^2)\,.
 \lb{2-2}
 \ee
Using (\ref{GK}) or Fourier transformed (\ref{G}) we find that
\be
G(p^2)=1/F(p^2)~.
\lb{FP}
\ee
In a Lorentz invariant theory described by equation (\ref{F}) Green's function (\ref{2-2}) is a function of the space-time interval $\sigma=(X-Y)^2$.
On the other hand, any given Green's function $G(\sigma)$  can be Fourier decomposed as in (\ref{2-2}).
Then using relation (\ref{FP}) we can restore the field equation this Green's function satisfies. In the coincidence limit, $X=Y$,
we find from (\ref{2-2})
\be
G(X,X)={2\o \Gamma({d\o 2})}{1\o (4\pi)^{d\o 2}}\int_0^\infty dp~p^{d-1}G(p^2)~~.
\lb{XY}
\ee
This limit is finite if function $G(p^2)$ is decaying faster than $1/p^d$  for large $p$.

As was shown in \cite{Nesterov:2010yi} in a theory described by the field equation (\ref{F}) entanglement entropy takes the form (for simplicity we take $\Sigma$ to be a $(d-2)$-dimensional plane)
\be
  S=\frac{A(\Sigma)}{12\cdot (4\pi)^{(d{-}2)/2}}
  \int_{\epsilon^2}^\infty {ds\o s}P_{d-2}(s)\,,
 \lb{Sent}
 \ee
where function $P_n(s)$ is defined as follows
\be
P_n(s)={2\o \Gamma({n\o 2})}\int_0^\infty dp~p^{n-1}~e^{-sF(p^2)}~~.
\lb{P}
\ee
Clearly, the function $P_{d-2}(s)$ is divergent in the limit $s\rightarrow 0$ for any function $F(p^2)$. This can be seen by taking literally this limit in the integral (\ref{P}) the integration over $p$ then is divergent in the upper limit. For small but finite $s$ this divergence is translated into divergence in variable $s$. Since arbitrary $F(p^2)$ means arbitrary Green's functions (\ref{2-2}), including those which are regular at $X=Y$ (i.e. integral in  (\ref{XY}) is finite),
we conclude that no matter how well  Green's function behaves in the coincidence limit the entanglement entropy remains UV divergent.
This is a slightly different formulation of the statement made in \cite{Nesterov:2010yi}.

\section{Padmanabhan's example}
\setcounter{equation}0
For simplicity in this section we consider the case of space-time dimension $d=4$. Green's function (\ref{3}) gives us an example of Green's function which is  regular in coincidence limit $X=Y$. As was noted in a recent paper
\cite{Padmanabhan:2010wg} function (\ref{3}) can be represented in a form similar to (\ref{GK})
\be
&&G_L(X,Y)=\int_0^\infty ds\ H(s,X,Y)~~,\nonumber \\
&&H(s,X,Y)={1\o (4\pi s)^{2}} e^{-|X-Y|^2/4s-L^2/4s}~~.
\lb{H}
\ee
If function $H(s,X,Y)$ was a heat kernel  then the $L$-dependent term in (\ref{H}) would regularize all UV divergences so that the effective action (\ref{W}) and entanglement entropy
(\ref{Sent}) would appear UV finite. This is the point of view advocated in \cite{Padmanabhan:2010wg}. However, despite the apparent similarity
of (\ref{H}) and (\ref{GK}) the function $H(s,X,Y)$ is NOT a heat kernel: it does not satisfy neither the field equation (\ref{K}) for any differential operator $\cal D$ nor the ``initial condition'' at $s=0$, i.e. it does not reproduce a delta-function.

On the other hand, the Fourier transform for Green's function $G_L(X,Y)$ (\ref{3}) is well defined.  We find that \cite{Padmanabhan:1996ap}
\be
G(p^2)={L\o p}\ K_1(pL)~~.
\lb{GP}
\ee
The corresponding heat kernel is defined by (\ref{2-K}) taking into account relation (\ref{FP}).
Hence we can reconstruct the relevant field equation  $F(\Box)\phi=0$ and obtain
\be
F({\Box})={\sqrt{\Box}\o L}\ {1\o K_1(L\sqrt{\Box})}~~.
\lb{FP2}
\ee
For small values of $p$ function $F(p^2)=G^{-1}(p^2)$  behaves as $p^2$ while for large $p$ it grows exponentially
\be
F(p^2)\simeq \sqrt{2\o \pi L}\ p^{3/2}\ e^{pL}~~.
\lb{FL}
\ee
For  a function  $F(p^2)$ with  asymptotic behavior (\ref{FL}) we have that
\be
P_2(s)={1\o L^2}\ (\ln{s\o L^2})^2+O\left(\ln{s\o L^2}\right)~~.
\lb{Pas}
\ee
Hence  entanglement entropy (\ref{Sent}) is UV divergent and the leading divergence
\be
S={A\o 48\pi L^2}\ {1\o 3}(\ln{\epsilon^2\o L^2})^3~~
\lb{Sdiv}
\ee
is logarithmic. This is despite the fact that Green's function (\ref{3}) is completely regular in the coincidence limit.

\section{Conclusions}
\setcounter{equation}0
In this note we have emphasized that, perhaps contrary to intuition, the short distance regularity of Green's function does not  imply the UV finiteness of  entanglement entropy. In fact our statement is rather general: in any theory characterized by   Lorentz invariant Green's
function the corresponding entanglement entropy is UV divergent, the degree of divergence is determined by the behavior of Fourier transform $G(p^2)$ (propagator) for large values of $p$. The requirement of the Lorentz symmetry is not essential. As we show in  \cite{Nesterov:2010yi}
entanglement entropy  remains UV divergent even if the Lorentz symmetry is violated by a generic term in the field operator.

Comparing our results with approach of Padmanabhan \cite{Padmanabhan:2010wg} we should note that from the field theoretical point of view his prescription to use the function $H(s,X,Y)$ instead of heat kernel can be viewed as a new method of regularization of the UV divergences similar to the Pauli-Villars regularization.
In his regularization  parameter $L$ plays the role of the UV cut-off similar to the parameter $\epsilon$ in the proper time regularization which we use. Then the divergences  in $\epsilon$ of effective action or entropy  are just replaced by equivalent divergences when $L$ is taken to zero.

\bigskip

The postdoctoral position of D.N. is financed by the University of Tours. D.N. was also supported by the RFBR grant 08-02-00725.


\end{document}